\documentclass[aps,superscriptaddress,showpacs]{revtex4}
\usepackage{amssymb,epsfig,latexsym}
\usepackage{amsmath,amscd}
\usepackage{graphicx}
\usepackage[normalem]{ulem}
\usepackage[dvips]{color}

\def\half{{\textstyle{1\over2}}}
\begin{document}

\title{Sensitivity of the neutron star r-mode instability window to the density dependence of the nuclear symmetry energy}

\author{De-Hua~Wen}
\affiliation{Department of Physics and Astronomy, Texas A\&M University-Commerce, Commerce, Texas 75429-3011, USA}
\affiliation{Department of Physics, South China University of Technology,Guangzhou 510641, P.R. China}
\author{W.~G.~Newton}
\affiliation{Department of Physics and Astronomy, Texas A\&M University-Commerce, Commerce, Texas 75429-3011, USA}
\author{Bao-An~Li}
\affiliation{Department of Physics and Astronomy, Texas A\&M University-Commerce, Commerce, Texas 75429-3011, USA}
\date{\today}


\begin{abstract}
Using a simple model of a neutron star with a perfectly rigid crust constructed with a set of crust and core equations
of state that span the range of nuclear experimental uncertainty in the symmetry energy, we calculate the instability window for the onset of the  Chandrasekhar-Friedmann-Schutz (CFS) instability in r-mode oscillations for canonical neutron stars ($1.4 M_{\odot}$) and massive neutron stars ($2.0 M_{\odot}$). In these models the crust-core transition density, and thus crustal thickness, is calculated consistently with the core equation of state (EOS). The EOSs are calculated using a simple model for the energy density of nuclear matter and probe the dependence on the symmetry energy by varying the slope of the symmetry energy at saturation density $L$ from 25 MeV (soft symmetry energy and EOS) to 115 MeV (stiff symmetry energy and EOS) while keeping the EOS of symmetric nuclear matter fixed. For the canonical neutron star, the lower bound of the r-mode instability window is reduced in frequency by $\approx150$ Hz from the softest to the stiffest symmetry energy used, independent of mass and temperature. The instability window also drops by $\approx 100$ Hz independent of EOS when the mass is raised from $1.4 M_{\odot}$ to $2.0 M_{\odot}$. Where temperature estimates are available, the observed neutron stars in low mass X-ray binaries (LMXBs) have frequencies below the instability window for the $1.4 M_{\odot}$ models, while some LMXBs fall within the instability window for $2.0 M_{\odot}$ stars if the symmetry energy is relatively stiff, indicating that a softer symmetry energy is more consistent with observations within this model. Thus we conclude that smaller values of $L$ help stabilize neutron stars against runaway r-mode oscillations. The critical temperature, below which no star can reach the instability window without exceeding its Kepler frequency, varies by nearly an order of magnitude from soft to stiff symmetry energies. When the crust thickness and core EOS are treated consistently, a thicker crust corresponds to a lower critical temperature, the opposite result to previous studies in which the transition density was independent of the core EOS.
\end{abstract}


\maketitle

\section{Introduction}

It has long been recognized that the Chandrasekhar-Friedmann-Schutz (CFS) instability \cite{Chandrasekhar70,Friedmann78} of oscillation modes may play an important role in generating detectable gravitational radiation from neutron stars, potentially providing a probe of the interior structure of neutron stars \cite{Lindblom98,Owen98, Andersson01,Andersson11,Kokkotas11}. Among the multitudinous oscillation modes, the r-mode, a class of purely axial inertial modes whose restoring force is the Coriolis force, is regarded as particularly interesting. Its CFS instability can spin down newborn neutron stars to the observed slower periods of young pulsars  \cite{Lindblom98,Owen98, Andersson01} ( $\Omega/\Omega_{k}<0.1$, where $\Omega_{k}$ is the break-up (Kepler) angular frequency). Here we focus on the possibility that the r-mode instability can limit the spin-up of accretion powered millisecond (ms) pulsars in low mass X-ray binaries (LMXBs). If accretion from the neutron star's companion continues for long enough unopposed, it should be able to spin up the star to its Kepler frequency. However, the observed population of ms pulsars all have rotation frequencies a factor of at least two below their theoretical upper limit, even within the uncertainties of the neutron star equation of state (EOS). Above the frequency at which the CFS instability sets in, the gravitational waves generated may be sufficient to prevent the neutron star from spinning up further by radiating away angular momentum \cite{Bildsten1998,Andersson1999}. For this explanation of the observed frequency cut-off to be plausible, the r-mode instability window, defined as the frequency above which the CFS instability is triggered for r-modes, should be below the Kepler frequency of the neutron stars, but above the highest observed frequency. The damping mechanisms that counter the CFS instability are temperature dependent, and therefore so is the critical frequency for the onset of the instability. The region in frequency-temperature space above the critical frequency is referred to as the instability window.

Significant progress has been made in studying oscillations in neutron stars with the inclusion of realistic physics \cite{Andersson11,Kokkotas11}. Particularly, if one assumes heating of the star due to accretion is balanced by neutrino cooling, and that the angular momentum gains from the spin-up torque are balanced by the loss of angular momentum due to the gravitational radiation from the unstable r-mode, Ho $et ~al.$ find that some of the neutron stars in LMXBs are located in the instability region \cite{Andersson11}. This indicates an incomplete understanding of the complex physics of the growth and dissipation of the instability. These new results suggest we need to check our understanding of the instability of the oscillation modes in the light of new observations and, from our perspective, new progress in nuclear theory.

Most previous studies of the r-mode instability focused on a canonical neutron star described by a polytropic EOS with fixed mass and radius (e.g. $M=1.4 M_{\odot}$ and $R=12.5 ~\textrm{km}$~\cite{Lindblom98,Owen98,Andersson01}). The recent discovery of a neutron star with a mass $M=2 M_{\odot}$ \cite{Demo10} reminds us that we should also study the r-mode instability windows of more massive neutron stars.

In old neutron stars, the presence of a solid crust plays an important role in the balance between the growth of the r-mode amplitude due to the CFS instability and its dissipation \cite{Bildsten00,Andersson00,Lindblom00,Rieutord01,Peralta06,Glampedakis06}. One simple, but useful, model assumes a perfectly rigid crust \cite{Bildsten00,Andersson00,Lindblom00}. This  provides an upper limit on the instability window because the viscous boundary layer (VBL) between the fluid core and crust is then maximally dissipative. A real crust is expected to be elastic, and the oscillation of the core could partially penetrate into the crust, which would decrease the dissipation from the core-crust boundary layer and therefore widen the instability window \cite{Peralta06,Glampedakis06,Andersson11}. It has been shown that the dissipation in the VBL between the crust and core is sensitive to the crust thickness \cite{Lindblom00}, and when one relaxes the assumption of perfect rigidity, the penetration of the mode into the elastic crust is also sensitive to the crust thickness \cite{Levin2001}.

The crust thickness depends on the radius of the star (and thus the core EOS) and the crust-core transition density (and thus the crust EOS). In investigations of the r-mode instability to date, the crustal thickness has not been calculated consistently with the core EOS; usually the crust-core transition density has been taken to be a canonical value $\sim1.5\times10^{14}$ g$\cdot$ cm$^{-3}$, or been allowed to vary independently of the core EOS \cite{Lindblom00}. This is mainly because efforts have been focused on the highly challenging task of developing the necessary formalism and computational tools to study the fluid dynamics of the r-mode. Meanwhile, over the past decade, the physics of the crust-core transition has been well explored and correlations with measurable quantities in nuclear structure and reactions established \cite{Ruster06,Chamel08,Xu09,Ducoin11}. Although more work is needed to study the complex physics in this regime, it is interesting to reinvestigate the effect of the crust on the r-mode CFS instability by employing the latest developments in our knowledge of the crust. Particularly, it is useful to establish quantitatively the effect of the crust thickness on the r-mode instability window and compare with other sources of uncertainty in modeling r-modes.

The main uncertainties in the crust-core transition arise from the uncertainties in our knowledge about the density dependence of the nuclear symmetry energy \cite{Lattimer04}. Significant progress has been made in constraining the density dependence of the symmetry energy in recent years, see, e.g.,~\cite{Li2008,Xiao09,Tsang09,Cen09,Leh09}; these experimental constraints lead to a range of crust-core transition densities and stiffnesses of core EOSs which we will explore in this paper.

In section II we review the formalism used to calculate the r-mode instability window for neutron stars with a rigid crust. In section III we describe the crust and core EOSs used and the resulting transition densities. In section IV we present the results before concluding in section V.

\section{Stability of the r-mode in a rigid crust neutron star}

According to the work of Lindblom, Owen, $et ~al$, the
timescale for the gravitational radiation driven growth of the r-mode instability for a
neutron star with rigid crust can be calculated by
\cite{Lindblom98,Owen98,Lindblom00}
\begin{equation}\label{TGR}
 {1\over\tau_{GR}} =  {32\pi G \Omega^{2l+2}\over c^{2l+3}}
{(l-1)^{2l}\over [(2l+1)!!]^2}  \times\left({l+2\over
l+1}\right)^{2l+2} \int_0^{R_c}\rho r^{2l+2} dr,
\end{equation}
and  the damping timescale for the r-mode due to viscous dissipation at the boundary layer of the perfectly rigid crust and fluid core can be
evaluated by \cite{Lindblom00}
\begin{equation}\label{TV}
 \tau_v  = \frac{1}{2\Omega} \frac{{2^{l+3/2}(l+1)!}}{l(2l+1)!!{\cal
 I}_l}
 \times \sqrt{2\Omega
R_c^2\rho_c\over\eta_c} \int_0^{R_c}
{\rho\over\rho_c}\left({r\over R_c}\right)^{2l+2} {dr\over R_c},
\end{equation}
where   $R_c$, $\rho_c$ and $\eta_c$ are the radius, density, and
the viscosity of the fluid at the core-crust interface. Here we
only consider the case $l=2$,  with ${\cal I}_2 =
0.80411$~\cite{Lindblom00,Rieutord01}.
As the temperature decreases below about
$10^9$~K,  the neutron star shear viscosity is
dominated by the electron-electron scattering, which has a density
and temperature dependence \cite{Flowers79,Cutler87}
\begin{equation}
 \eta_{ee}=6.0\times 10^6 \rho^2 T^{-2}~ (g\cdot cm^{-1}\cdot
 s^{-1}).
\end{equation}
And for the temperatures above about $10^{9}$K, it is expected
that neutron-neutron scattering becomes the dominant dissipation
mechanism, and its viscosity is given by \cite{Flowers79,Cutler87}
\begin{equation}
\eta_{nn}=347\rho^{9/4} T^{-2} ~ (g\cdot cm^{-1}\cdot s^{-1}).
\end{equation}

According to Eq. \ref{TGR}, the gravitational radiation time scale
$\tau_{GR}$ has a spin-frequency dependence $ \sim \Omega^{-6}$,
so it is convenient to define a fiducial time scale
$\tilde{\tau}_{GR}$ like
\begin{equation}\label{TGR1}
 \tau_{GR} = \tilde{\tau}_{GR}
\left({\Omega_o\over\Omega}\right)^6,
\end{equation}
and thus $\tilde{\tau}_{GR}$ is the gravitational radiation time
scale at $\Omega_o=\sqrt{\frac{3 G M}{4R^{3}} }$. Similarly,
according to Eq. \ref{TV}, it is also advantageous to define a
fiducial viscous time scale $\tilde{\tau}_{v}$ as
\begin{equation}\label{TV1}
 \tau_v = \tilde{\tau}_v \times
 T_{8}\times\left({\Omega_o\over\Omega}\right)^{1/2},
\end{equation}
where $T_{8}=T/10^{8}K$.

The critical rotation frequency $\Omega_{c}$ is defined as
the frequency at which the energy dissipation rate due to the viscosity
of the boundary layer is exactly balanced by the rate of energy gain by r-mode as its amplitude grows.
This definition is equivalent to $\tau_v=\tau_{GR}$. According to
Eqs. \ref{TGR1} and \ref{TV1}, the critical rotation frequency can
be evaluated by
\begin{equation}\label{WC}
 {\Omega_c}
= \left({\tilde{\tau}_{GR}\over\tilde{\tau}_v}\right)^{2/11}
\times T_{8}^{2/11}\times \Omega_o.
\end{equation}
This is the  equation which determines the critical frequency, versus temperature, above which the CFS instability dominates the angular momentum evolution of the star - i.e. the r-mode instability window for a rigid crust neutron star. High temperatures $T>10^{10}K$ characteristic of a newborn neutron star are not considered here, at
which the bulk viscosity will be the dominant dissipation mechanism.

The Kepler frequency is the dynamical upper limit  on the spin frequency of a neutron star, so one can define a critical temperature below which the instability will be completely suppressed based on the Kepler frequency $\Omega_{k}\approx \frac{2}{3}\Omega_{0}$. According to Eq.\ref{WC}, it is easy to obtain
\begin{equation}
{T_c}  \approx 10^{8}\times \left({3\over
2}\right)^{11/2}\times{\tilde{\tau}_{GR}\over\tilde{\tau}_v}.
\end{equation}

The core EOS and crust-core transition densities enter into integrals in Eqs.\ref{TGR} and \ref{TV} through the density - radius relation $\rho(R)$ and the radius at which the crust-core transition occurs $R_{\rm c}$.

\begin{table}
\caption{\label{tab:table1} The transition densities $\rho_{c}$ of
the five selected slope parameters L. }
\begin{ruledtabular}
\begin{tabular}{cccccc}
L (MeV) & 25 & 45 & 65 & 85 & 105\\
\hline
$\rho_{c}$ $(fm^{-3})$ & 0.0940 & 0.0831 & 0.0762 & 0.0726 & 0.0703\\
 \end{tabular}
\end{ruledtabular}
\end{table}

\begin{figure}
\centering
\includegraphics [width=0.8\textwidth]{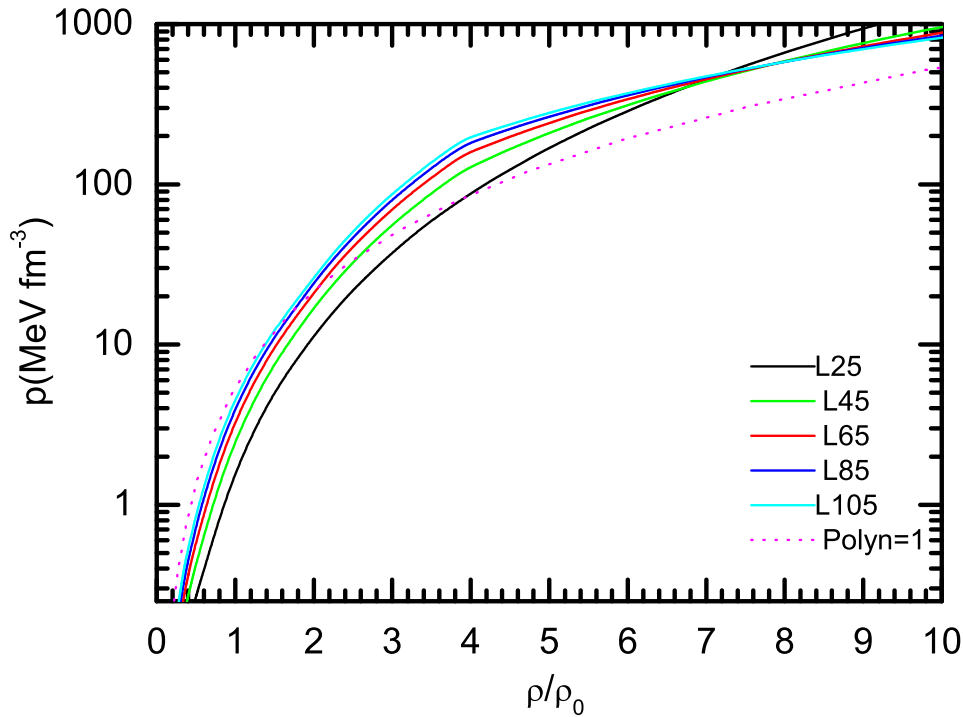}
\caption{\label{fig1} (Color online) The EOS of neutron-star matter versus the reduced baryon number density 
$\rho/\rho_0$. For comparison, a purely polytropic EOS $P=K \rho^{2}$ is
employed with parameter $K$ chosen so that a $1.4 M_{\odot}$ model has a radius 12.53
km (marked as Polyn=1).}
\end{figure}
\section{Crust and core equations of state}

The energy per particle of an idealized system of matter at a uniform density $\rho$, made up of neutrons and protons, in the absence of Coulomb interactions, is referred to as the uniform nuclear matter EOS $E(\rho,\delta)$. Here $\delta$ is a measure of the isospin asymmetry of the matter: if $x$ is the fraction of protons, $\delta = 1-2x$. $E(\rho,\delta)$ can be usefully decomposed into the two extremes of isospin asymmetry: symmetric nuclear matter (SNM) $E_{\rm SNM}$ = $E(\rho,\delta=0)$ and pure neutron matter (PNM) $E_{\rm PNM}$ = $E(\rho,\delta=1)$ via the introduction of the symmetry energy. The symmetry energy is defined $S(n) = \half \partial^2 E(n,\delta)/\partial \delta^2_{\delta=0}$, and approximately $E_{\rm PNM}(n) \approx E_{\rm 0}(n) + S(n)$. Most importantly, the slope of the symmetry energy $L$ at nuclear saturation density $n \approx 0.16$ fm$^{-3}$ is strongly correlated with the crust-core transition density and the pressure inside a neutron star at densities just above and below saturation; it is defined $L = \partial S(n) / \partial \chi |_{n = n_{\rm s}}$ where $\chi = \frac{n-n_{\rm s}}{3n_{\rm s}}$. In all models considered here, the EOS of symmetric nuclear matter $E_{\rm SNM}$ is fixed with an incompressibility modulus of $K_0$ = 240 MeV. The change in stiffness of the EOS for neutron-star matter is completely due to varying the symmetry energy slope parameter $L$.

\begin{figure}
\centering
\includegraphics [width=0.8\textwidth]{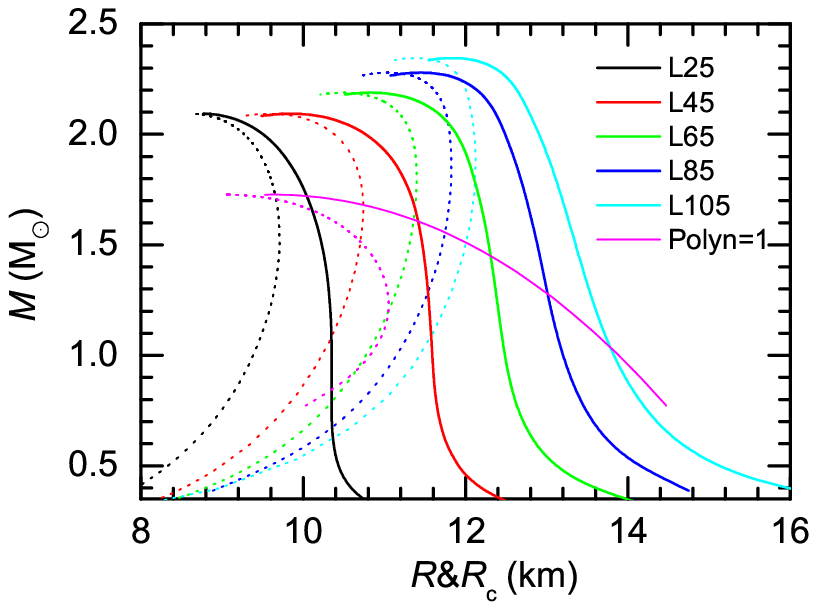}
\caption{\label{fig2} The mass-radius relations
(solid lines) and the radii $R_{c}$ of the core (dot lines) for our selected EOSs,
where the core-crust transition densities are consistent with the slope of the symmetry energy used (see Tab. \ref{tab:table1}), and the
transition density of polytropic sequence is chosen as 1.5 $\times
10^{14}$ g$\cdot$cm$^{-3}$.}
\end{figure}

The crust-core transition densities used in this work are obtained from a simple compressible liquid drop model (CLDM) for the crust \cite{Newton2011}. The nuclear matter contribution to the crust EOS and the EOS of beta-equilibrated matter in the core is obtained from the Modified Skyrme-Like (MSL) model \cite{MSL01}. The MSL interaction contains parameters directly related to the properties of nuclear matter at saturation density, such as the slope of the symmetry energy $L$, allowing them to be independently adjusted and giving us the ability to scan over the experimentally constrained values. We take as a conservative range of $L$ from model analyses of terrestrial nuclear laboratory experiments $25 < L < 115 $ MeV \cite{Chen2005,Li2005,Shetty2007,Klimkiewicz2007, Danielewicz2007,Li2008,Cen09,Tsang09,LieWenChen2010,ChangXu2010}, although it is worth noting that the most recent constraints \cite{Cen09,LieWenChen2010,ChangXu2010,Liu2010,RocaMaza2011}, coupled with inference from theoretical calculations of pure neutron matter \cite{Hebeler2010,Gandolfi2011}, place the value of $L$ in the lower half of this range 25 - 75 MeV. For each value of $L$, the magnitude of the symmetry energy at normal density is adjusted so that the EOS fits the most recent calculations of the EOS of PNM \cite{Hebeler2010,Gandolfi2011} as detailed in \cite{Newton2011}. The crust-core transition density is obtained by finding the density at which uniform neutron star matter becomes energetically favorable to inhomogeneous matter as calculated in the CLDM. The core EOS then continues on from the crust EOS at the transition density, calculated using the same MSL model. At high densities $\gtrsim 3 n_{\rm s}$, we expect the description of the matter in terms of just neutron and protons to break down (i.e. MSL becomes an invalid model), although the exact composition of the inner core is still uncertain. As a stand-in for our lack of knowledge in this regime, and also to ensure that our complete EOS is always sufficiently stiff to produce 2$M_{\odot}$ neutron stars as demanded by observations \cite{Demo10}, we smoothly join the MSL EOS to two polytropic EOSs of the form $P = K \epsilon^{(1+1/n)}$ in a similar way to \cite{Steiner2010}. The joins are made at energy densities of 300 MeV fm$^{-3}$ and 600 MeV fm$^{-3}$ by adjusting the constant $K$ to keep the pressure continuous at the join. The lower density polytrope has an index set at $n=0.5$, while the second index takes  a range $n=0.5-1.5$ for values of $L$ from 25-115 MeV respectively. Note that although the additions of the polytropes substantially changes the maximum neutron star mass, especially for small (soft) values of $L$, it does not substantially affect the radius and crust thickness of a neutron star of a given mass, as such properties are determined dominantly by the pressure of the star in the range $\gtrsim 1-2 n_{\rm s}$ through the symmetry energy slope parameter $L$ \cite{Lattimer04}. The EOSs are displayed in Fig. \ref{fig1} with the corresponding crust-core transition densities tabulated in Tab. \ref{tab:table1}. Interested readers can obtain numerical values of the EOSs used here from our website \cite{Will11}.

In order to compare with previous studies, we will also use a core EOS constructed entirely from a polytrope of index 1, a constant $K$ chosen to give a radius $R = 12.53$ km for a canonical neutron star, and with an independently variable crust-core transition density.

The mass-radius relation and the radius of the core are displayed in Fig. \ref{fig2}. A larger value of the slope parameter $L$ corresponds to a smaller value of a crust-core transition density, thus reducing the crust thickness; this is offset, however, by the increased radius of the star as the core EOS is stiffer at lower densities $1-2n_{\rm s}$, so that overall a larger slope
parameter $L$ corresponds to a thicker crust.

\section{Results and discussion}

\begin{figure}
\centering
\includegraphics [width=0.45\textwidth,height=8cm]{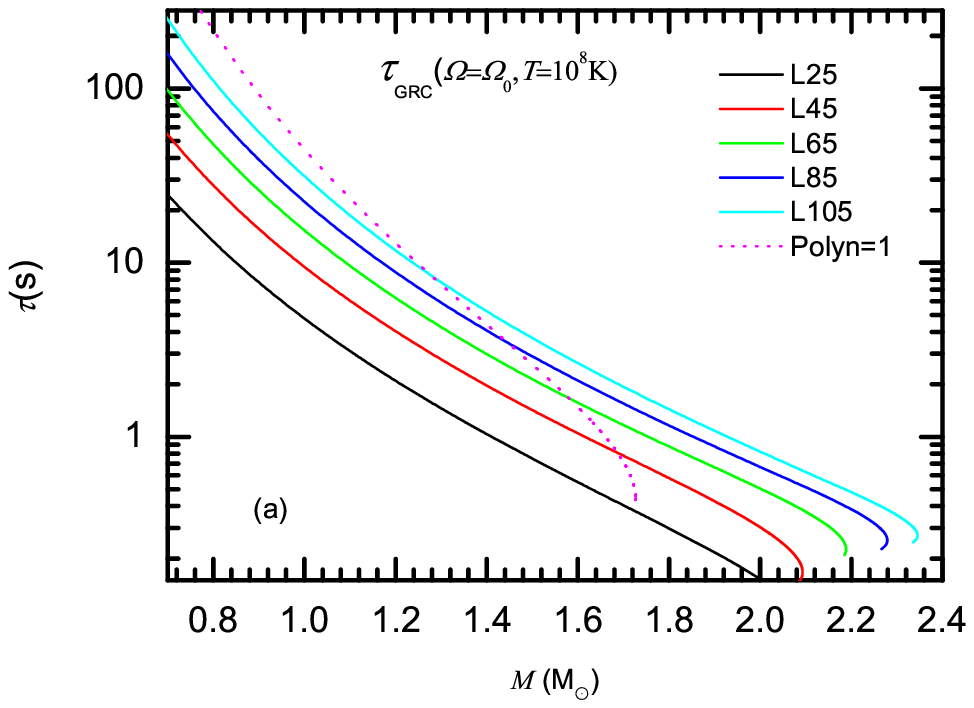}
\includegraphics [width=0.45\textwidth,height=8cm]{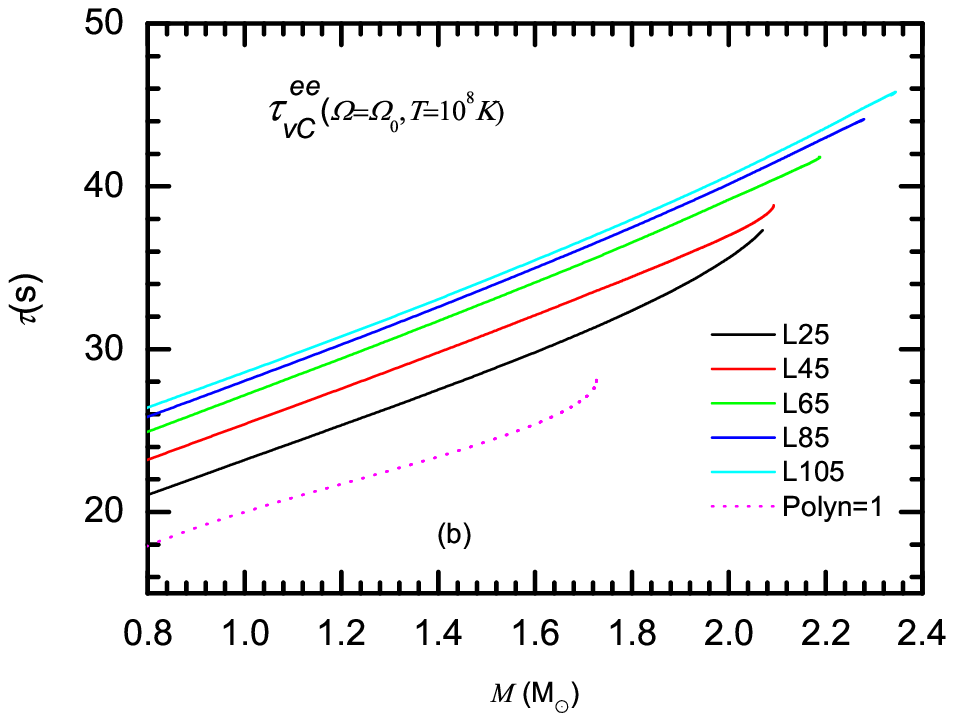}
\caption{\label{3Tgrc} (a) The time scales for the gravitational radiation driven r-mode instability for a neutron star with rigid crust as a function of the stellar masses, where $\Omega=\Omega_o=\sqrt{{3 G M}/({4R^{3}}) }$ and $T=10^{8}$ K. (b) The corresponding viscous dissipation time scales due to the electron-electron scattering at the core-crust boundary layer.}
\end{figure}

In this section we present our numerical results. Shown in Fig. \ref{3Tgrc} are the time scales of
the gravitational radiation (left window) and the shear viscosity (right window) which dissipates the r-mode
instability, respectively, as a function of the stellar masses. Here the neutron star is modeled with a rigid crust 
for the five adopted EOSs with $L$=25-105 MeV (which henceforth we shall label $L25, L45$ etc). The shear viscosity 
of the boundary layer is taken to be dominated by electron-electron scattering (EES). There is an evident mass 
dependence for the time scales of the gravitational radiation with the more massive neutron stars having a shorter timescale for angular momentum dissipation due to gravitational radiation. On the other hand, the time scales of the shear viscosity in the core-crust boundary layer dominated by EES increase with the stellar masses. According to Eq. \ref{WC}, one can see that there should be a distinct effect of the stellar
mass on the r-mode instability window.

\begin{figure}
\includegraphics [width=0.45\textwidth,height=8cm]{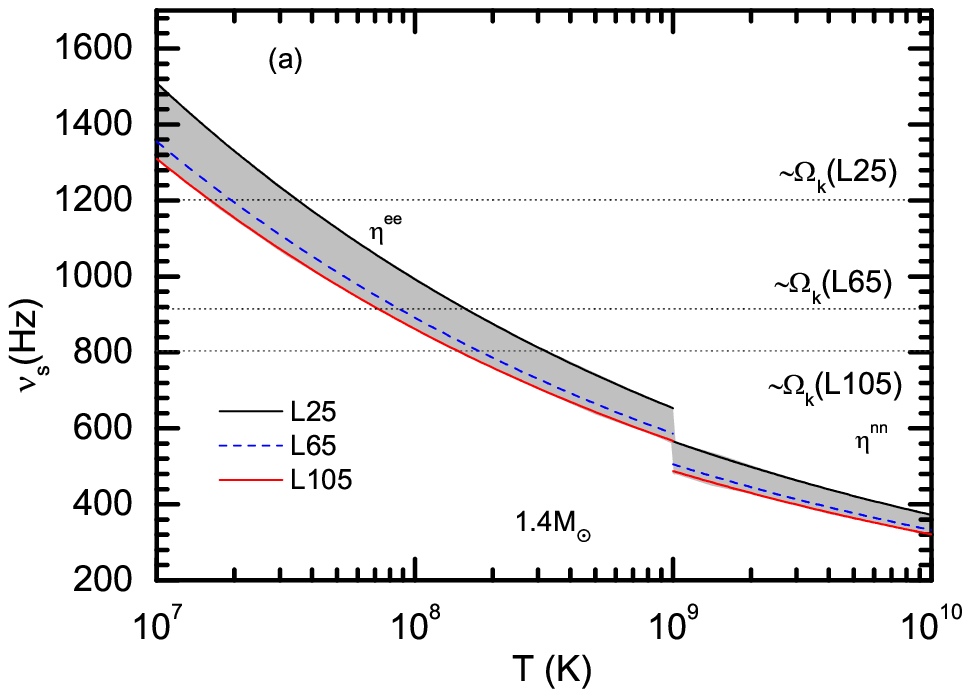}
\includegraphics [width=0.45\textwidth,height=8cm]{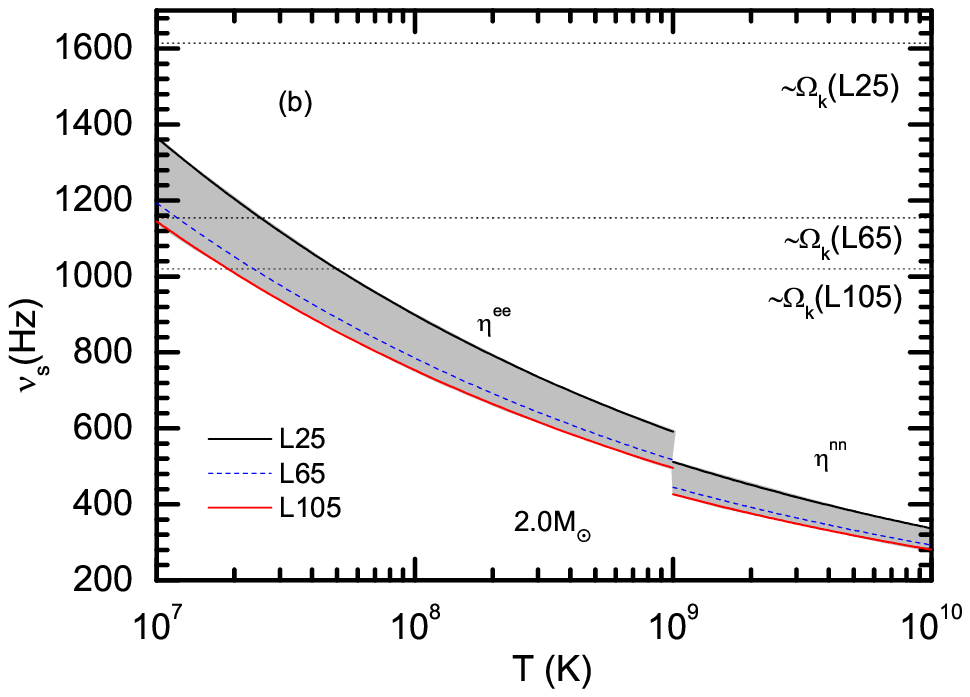}
\caption{\label{2ns} The lower boundary of the r-mode instability window for a $1.4 M_{\odot}$ (a) and $2.0 M_{\odot}$ (b) neutron star over the range of the slope of the symmetry energy $L$ consistent with experiment. Examples of the corresponding break-up (Kepler) frequencies are indicated by the dotted lines.}
\end{figure}
The r-mode instability windows for a neutron star of $1.4 M_{\odot}$ and $2.0 M_{\odot}$ with five selected slope parameters
are displayed in the left and right window of Fig. \ref{2ns}, respectively. Considering the values of slope parameter $L$ in the range between 25-105 MeV are consistent with the terrestrial nuclear laboratory data, we conclude that the boundary of the r-mode instability window for
a $M=1.4 M_{\odot}$ neutron star should be constrained in the shaded regions. It is worth noting that the r-mode instability window of the $1.4 M_{\odot}$ neutron star based on the standard $n=1$ polytrope with a radius of 12.53 km nearly superposes that of $L25$, at the lowest end of the experimentally constrained values. The discoveries of massive neutron stars, such as PRS J1614-2230
with  $M=1.97\pm 0.04 M_{\odot}$ \cite{Demo10} and EXO 0748-676 with $M\geq 2.10\pm 0.28 M_{\odot}$ km
\cite{Ozel06} reminds us to study the r-mode instability of a massive neutron star. The instability window of a massive star ($M=2.0 M_{\odot}$) is displayed in right window of Fig. \ref{2ns}. One can see that the massive neutron star has a wider instability
window compared with the canonical neutron star ($M=1.4 M_{\odot}$). Similarly, the r-mode instability window of the massive neutron
star is constrained in the shaded region.

\begin{figure}
\includegraphics [width=0.45\textwidth,height=8cm]{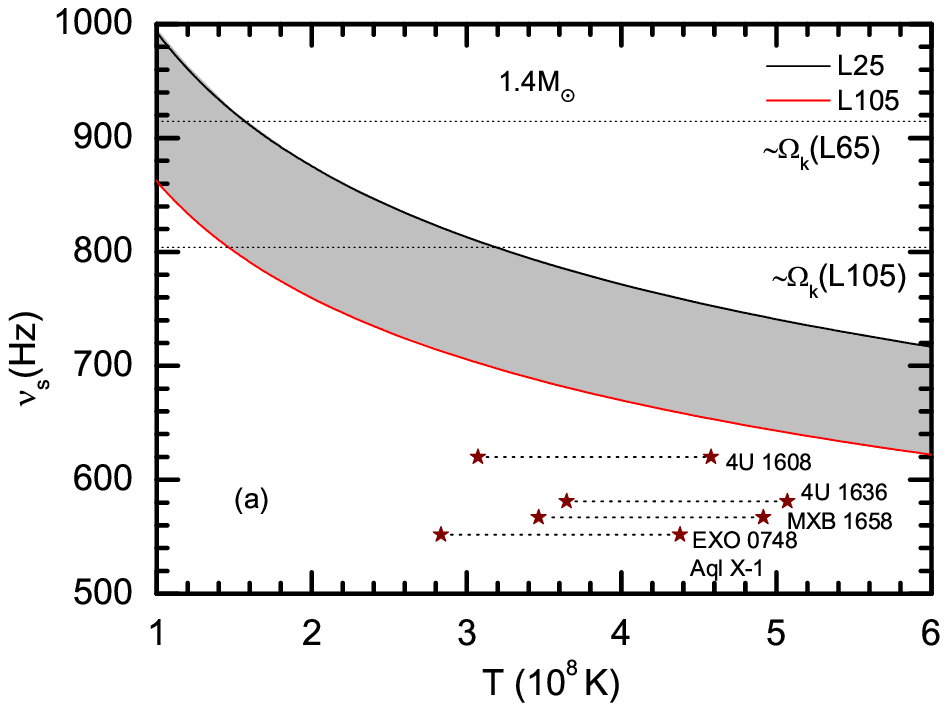}
\includegraphics [width=0.45\textwidth,height=8cm]{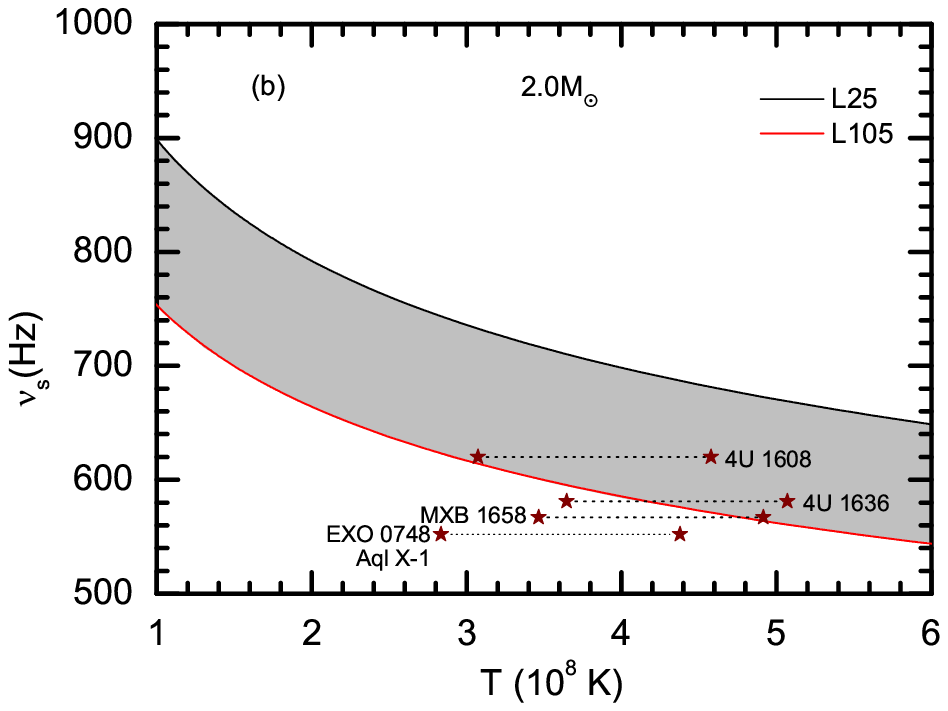}
\caption{\label{7windows1.4stars} The location of the
observed short recurrence time LMXBs \cite{Watts08, Keek10} in frequency-temperature space, together with the lower bound of the instability windows from Fig. 4 for a $1.4 M_{\odot}$ (a) and $2.0 M_{\odot}$ (b) neutron star. The core temperature $T$ of the LMXBs are
derived from their observed accretion luminosity and assuming the
cooling is dominant by the modified Urca neutrino emission process
for normal nucleons ($left ~stars$) or by the modified Urca
neutrino emission process for neutrons being superfluid and
protons being superconducting ($right ~ stars$) in the core
matter, respectively \cite{Andersson11}. Note that the
position of EXO 0748 almost superposes that of Apl X-1. }
\end{figure}

\begin{figure}
\centering
\includegraphics [width=0.7\textwidth]{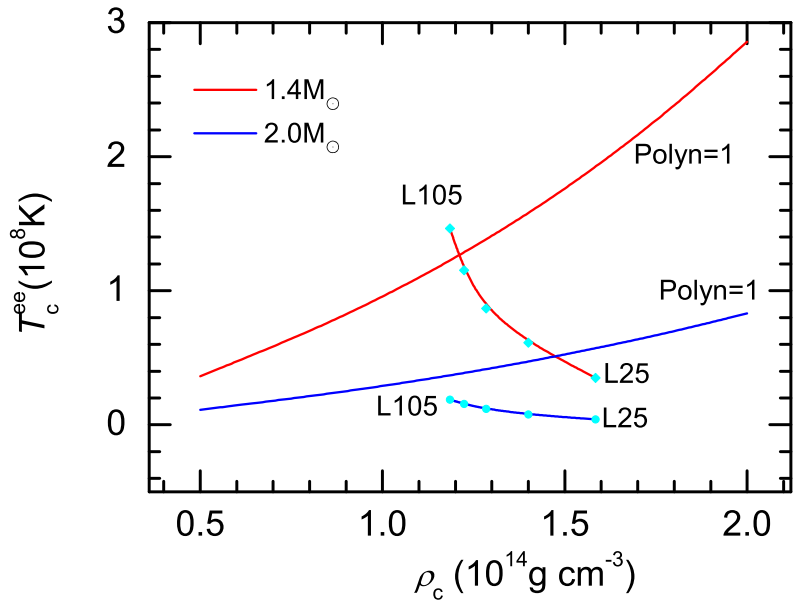}
\caption{\label{9temp} The critical temperature
$T_{c}$ for the onset of the CFS instability versus the crust-core transition densities (Tab.~1) over the range of the slope of the symmetry energy $L$ consistent with experiment for $1.4 M_{\odot}$ and $2.0 M_{\odot}$ stars. The transition density dependence of $T_{c}$ for the pure $n=1$ polytrope is also plotted.. The transition density is calculated consistently with the core EOS except for the pure polytrope, in which it is freely varied.}
\end{figure}
As the observed short recurrence time LMXBs have rapidly rotating
frequencies (such as 4U 1608-522 at 620Hz, 4U 1636-536 at 581Hz,
MXB 1658-298 at 567Hz, EXO 0748-676 at 552Hz and Apl X-1 at 549Hz)
\cite{Watts08, Keek10} and high core temperatures \cite{Andersson11},
we here focus on the location of these LMXBs in the r-mode instability
windows of neutron stars modeled
with rigid crusts. As the masses of the neutron stars in LMXBs are
not measured as accurately as that in the two-neutron-star
binaries, the positions of the LMXBs in the r-mode
instability windows are explored in both the canonical ($M=1.4
M_{\odot}$) and massive ($M=2.0 M_{\odot}$) neutron stars. The
results are shown in Fig. \ref{7windows1.4stars}, where the core temperature $T$ of the
LMXBs are derived from their observed accretion luminosity and
assuming the cooling is dominated by the modified Urca neutrino
emission process for normal nucleons ($left ~stars$) or by the
modified Urca neutrino emission process for neutrons being
superfluid and protons being superconducting ($right ~ stars$) in
the core \cite{Andersson11}. It is shown that for a $M=1.4
M_{\odot}$ neutron star, all the considered LMXBs lie outside of the
instability window, which is consistent with the finding that no
r-mode is currently excited in LMXBs because of the shortness of
the period of the r-mode activity  \cite{Levin99,Bondarescu07}.
For the massive neutron stars ( $M=2.0 M_{\odot}$), some of
the stars are in the constraint instability window. Considering
LMXBs should be out of the instability window \cite{Levin99,Bondarescu07}, from Fig. \ref{7windows1.4stars}, one
can conclude that either the masses of the neutron stars in LMXBs
are not so massive, or a small slope parameter $L$ is preferred.

The critical temperature ${T_c}$ is defined as the temperature below which the instability will be
completely suppressed even for neutron stars rotating at the Kepler frequency.  Shown in Fig. \ref{9temp} is
${T_c}$ versus the core-crust transition densities $\rho_{c}$ for both a $1.4 M_{\odot}$ and a $2.0 M_{\odot}$ star. The results for the purely polytropic EOS are displayed for comparison, in which case the transition density is a free parameter and is not varied accordingly. It is interesting to compare these results with those of Lindblom {\it et al.} \cite{Lindblom00} for a $1.4 M_{\odot}$ star. In that study, with a crust-core transition density of $1.5\times10^{14}$ g cm$^{-3}$, the lowest critical temperature obtained was $\approx 1.5 \times 10^8$K; in our study, at the same transition density, corresponding to $L=25$ MeV, and using a core EOS consistent with that value of $L$, the critical temperature is almost a factor of 3 lower, at $\approx 4\times10^7K$. Over the range of $L$ experimentally constrained, the critical temperature varies by a factor of almost 5 from $\approx 3\times10^7$K to $1.5\times10^8$K from soft to stiff EOSs, thus highlighting the importance of using consistent EOSs for both the core and crust. It is also important to note that the variation of the critical temperature has an opposite correlation with crust-core transition density when one treats it consistently with the core EOS as opposed to allowing the crust-core transition to vary for a fixed core EOS. Interestingly, for a $2.0 M_{\odot}$ star, the variation of ${T_c}$ with the transition density is also about a factor of 5 from $4\times10^6$K to $2\times10^7$K.

\section{Conclusions}

Within a simple model of a neutron star with a perfectly rigid crust we have calculated the r-mode instability window, with the dominant damping mechanism being viscosity due to electron-electron scattering at the crust-core boundary. Using a consistent description of the crust-core transition density and the core EOS we have explored the dependence of the instability window on the pressure of pure neutron matter through the symmetry energy slope parameter $L$ over the range of values constrained by terrestrial nuclear laboratory experiments and theoretical pure neutron matter calculations. The low frequency boundary of the instability window can vary by about 150Hz over this range for both the 1.4$M_{\odot}$ and 2.0$M_{\odot}$ neutron stars; however, the lower boundary is shifted down by $\approx 100$Hz for the 2.0$M_{\odot}$ star, placing some of the ms pulsars in LMXBs within the instability window for stiffer EOSs. Softer EOSs raise the lower frequency bound of the instability window; therefore, within this model, EOSs characterized by $L\lesssim 65$ MeV are more consistent with the observations, values of $L$ also currently more favored by recent nuclear experimental evidence and theoretical calculations \cite{Liu2010,RocaMaza2011,Hebeler2010,Gandolfi2011}.

The critical temperature above which the instability can set in at frequencies below the Kepler frequency is found to be in the range $3\times10^7$K - $1.5\times10^8$K for a 1.4$M_{\odot}$ star, and $4\times10^6$K - $2\times10^7$K  for a 2.0$M_{\odot}$ star. The uncertainty in the symmetry energy slope parameter $L$ leads to a factor of 5 difference in the critical temperature for both neutron stars considered. We notice that the critical temperature correlates inversely with the crust-core transition density in both cases. This is in contrast to the treatment of the transition density independently of the core EOS, in which the correlation is positive, as found in \cite{Lindblom00}. The crust thickness is correlated positively with the transition density and the radius of the star; the former increases with a softer EOS, while the latter decreases. Only when both are treated consistently does the correct dependence emerge.

It is well known that the model for the r-mode instability window presented here is much oversimplified; the crust is elastic and r-modes can penetrate some way into it, increasing the damping timescales and lowering the instability window. The crust-core boundary might be mediated by exotic phases of inhomogeneous nuclear matter (`pasta') whose mechanical properties are still yet to be accurately calculated. Both scenarios still depend sensitively on the crust thickness and core EOS. Our model ignores the dissipative mechanisms available to superfluid components in the core (e.g. mutual friction) as well as a host of other physical effects which can alter the position of the instability window \cite{Andersson11}, or prevent the growth of r-modes in the instability window to amplitudes at which they can effectively be damped \cite{Bondarescu07}. The extent to which any of these mechanisms is effective at altering the properties of unstable r-modes is still rather model dependent. However, we have shown that the consistent treatment of the EOSs for the crust and core can have a significant effect on the position of the instability window, with a softer symmetry energy producing greater consistency with the observations, and such a consistent treatment can be implemented relatively simply.

\section*{ACKNOWLEDGEMENTS}
We would like to thank Dr. Li Ou and Mr. M. Gearheart for helpful discussions and assistance, and the referee for helpful comments and suggestions. This work is supported in part by the National Aeronautics and Space Administration under grant NNX11AC41G issued through the Science Mission Directorate and the National Science Foundation under grants PHY-0757839 and PHY-1068022 and the Texas Coordinating Board of Higher Education under grant No. 003565-0004-2007. D.H. Wen is also supported in part by the National Natural Science Foundation of China under Grant No.10947023,  and the Fundamental Research Funds for the Central University, China under Grant No.2012zz0079. The project is sponsored by SRF for ROCS, SEM. We thank the referee for useful comments.

\end{document}